\documentclass[prl,twocolumn,tightenlines,preprintnumbers,floatfix,nofootinbib]{revtex4}
\usepackage{graphicx,amsfonts,amssymb,amsmath}
\usepackage{array}

\def\be{\begin{equation}}
\def\ee{\end{equation}}
\def\ba{\begin{eqnarray}}
\def\ea{\end{eqnarray}}
\def\ge{\mathrel{\raise.3ex\hbox{$>$\kern-.75em\lower1ex\hbox{$\sim$}}}}
\def\la{\mathrel{\raise.3ex\hbox{$<$\kern-.75em\lower1ex\hbox{$\sim$}}}}

\def\simgt{\mathrel{\raise.3ex\hbox{$>$\kern-.75em\lower1ex\hbox{$\sim$}}}}
\def\simlt{\mathrel{\raise.3ex\hbox{$<$\kern-.75em\lower1ex\hbox{$\sim$}}}}

\newcommand{\bi}[1]{\bibitem{#1}}
\newcommand{\fr}[2]{\frac{#1}{#2}}

\newcommand{\nc}{\newcommand}

\nc{\gone}{\bar g_{\pi NN}^{(1)}}
\nc{\gzero}{\bar g_{\pi NN}^{(0)}}
\nc{\al}{\alpha}
\nc{\ga}{\gamma}
\nc{\de}{\delta}
\nc{\ep}{\epsilon}
\nc{\ze}{\zeta}
\nc{\et}{\eta}

\nc{\Th}{\Theta}
\nc{\ka}{\kappa}
\nc{\rh}{\rho}
\nc{\si}{\sigma}
\nc{\ta}{\tau}
\nc{\up}{\upsilon}
\nc{\ph}{\phi}
\nc{\ch}{\chi}
\nc{\ps}{\psi}
\nc{\om}{\omega}
\nc{\Ga}{\Gamma}
\nc{\De}{\Delta}
\nc{\La}{\Lambda}
\nc{\Si}{\Sigma}
\nc{\Up}{\Upsilon}
\nc{\Ph}{\Phi}
\nc{\Ps}{\Psi}
\nc{\Om}{\Omega}
\nc{\ptl}{\partial}
\nc{\del}{\nabla}
\nc{\ov}{\overline}
\nc{\newcaption}[1]{\centerline{\parbox{15cm}{\caption{#1}}}}

\def\beq{\begin{equation}}
\def\eeq{\end{equation}}
\def\bmat{\begin{displaymath}}
\def\emat{\end{displaymath}}
\def\bear{\begin{eqnarray}}
\def\eear{\end{eqnarray}}
\def\bery{\begin{array}}
\def\ery{\end{array}}
\def\bit{\begin{itemize}}
\def\eit{\end{itemize}}
\def\ben{\begin{enumerate}}
\def\een{\end{enumerate}}
\def\btab{\begin{tabular}}
\def\etab{\end{tabular}}
\def\btbl{\begin{table}}
\def\etbl{\end{table}}
\def\bfig{\begin{figure}[htb]}
\def\efig{\end{figure}}
\def\bpic{\begin{picture}}
\def\epic{\end{picture}}

\def\ga{\mathrel{\raise.3ex\hbox{$>$\kern-.75em\lower1ex\hbox{$\sim$}}}}
\def\la{\mathrel{\raise.3ex\hbox{$<$\kern-.75em\lower1ex\hbox{$\sim$}}}}
\def\gappeq{\mathrel{\rlap {\raise.5ex\hbox{$>$}}
{\lower.5ex\hbox{$\sim$}}}}
\def\lappeq{\mathrel{\rlap{\raise.5ex\hbox{$<$}}
{\lower.5ex\hbox{$\sim$}}}}

\def\gyr{{\rm \, G\kern-0.125em yr}}
\def\mev{{\rm \, Me\kern-0.125em V}}
\def\gev{{\rm \, Ge\kern-0.125em V}}
\def\tev{{\rm \, Te\kern-0.125em V}}

\begin{document}

\setcounter{page}{1}

\title{New Parity-Violating Muonic Forces}

\author{
Brian Batell$^{\,(a)}$, David McKeen$^{\,(b)}$, and Maxim Pospelov$^{\,(a,b)}$}

\affiliation{$^{\,(a)}${\it Perimeter Institute for Theoretical Physics, Waterloo,
ON N2J 2W9, Canada}\\
$^{\,(b)}${\it Department of Physics and Astronomy, University of Victoria, 
     Victoria, BC V8P 1A1, Canada}}

\begin{abstract}

The recent discrepancy between proton charge radius measurements extracted 
from electron-proton versus muon-proton systems is suggestive 
of a new force that differentiates between lepton species. 
We 
identify a class of models with gauged right-handed muon number, which contains new vector and scalar force carriers
at the $\sim$100 MeV scale or lighter, that is consistent with observations.
Such forces 
would lead to an 
enhancement by 
several orders-of-magnitude of the parity-violating asymmetries 
in the scattering of low-energy muons on nuclei.
The relatively large size of such asymmetries, $O(10^{-4})$, opens up the  
possibility for new tests of parity violation in neutral currents with 
existing low-energy muon beams.

\end{abstract}

\maketitle

\newpage

{\em Introduction.} 
There has been much interest as of late in the possibility of new gauge forces existing in the MeV-GeV scale,
stimulated in part by the prospect of a light mediator between dark matter 
and the standard model (SM)
(see, {\em e.g.}, \cite{DM}). While many models of this type can 
be explored, a great deal of attention 
has been given to
a new $U(1)$ gauge boson
$V$ 
kinetically mixed with hypercharge \cite{Holdom} . At low energies $V$ appears as a massive copy of the ordinary photon,
\be
{\cal L} = -\fr{1}{4} V_{\mu\nu}^2 + \fr{1}{2} m_V^2 V_\mu^2 + \kappa V_\mu J_\mu^{EM},
\label{kinmix}
\ee 
where $\kappa$ is the mixing angle parameter. 
The conservation of the electromagnetic current and the 
absence of any intrinsic parity/flavor/CP violation in the 
interaction  of $V$ with the SM fermions can hide this force from 
very powerful symmetry tests. The model (\ref{kinmix}), while perhaps the simplest, is not the 
unique possibility for new gauge interactions below the weak scale \cite{Ringwald}. 

While the astroparticle physics incentives
are rather speculative, 
an additional motivation for a new light gauge boson is provided by terrestrial experiments.
Among several discrepant low-energy measurements, the recent determination of the 
proton charge radius using muonic hydrogen \cite{muH} and the long-standing measurement of the anomalous magnetic moment of the muon \cite{g-2}   
may be manifestations of a new sub-GeV scale force carrier 
that couples preferentially to muons. In this letter 
we argue that if such discrepancies are caused by a new muon-specific gauge force, 
one should expect parity non-conservation (PNC) in the scattering of muons on nuclei far above the SM level. We point out the feasibility of a 
dedicated search for the PNC asymmetry, enhanced 
to $O(10^{-4})$ level, with existing low-energy muon beams.

With our present 
understanding of the strong interactions, the charge radius of the proton $r_p$ 
cannot be computed from first principles but instead must 
be extracted from experiment. The comparison of 
$r_p$ values obtained using
different experimental methods provides a consistency check of
QED theory
and 
constrains a variety of new physics scenarios. 
Currently, there are 
three competitive ways of determining $r_p$: 1) high-precision measurements of the 
atomic levels in hydrogen and deuterium, 2) direct electron-proton scattering experiments, and
3) the measurement of the Lamb shift in muonic hydrogen. The
most precise determinations 
currently read \cite{CODATA,scattering,muH}
\begin{eqnarray}
r_{p,1} &=& 0.8768(69)~{\rm fm}~~~~~\,{\rm atomic~ H,D},\\
r_{p,2} &=& 0.879(8)~{\rm fm}~~~~~~~~\,e-p{\rm~scattering}, \\
r_{p,3} &=& 0.84184(67)~{\rm fm}~~~~{\rm muonic~ H}.
\end{eqnarray}
The 
$r_p$ values obtained from $e-p$ systems are consistent with each other and 
significantly differ from the $r_p$ value extracted from the muonic hydrogen Lamb shift,
\begin{eqnarray}
\label{pattern}
&r_{p,1}\simeq r_{p,2} > r_{p,3},&\\
&\Delta r^2 \equiv(r_p)^2_{e-p~{\rm results}} - (r_p)^2_{\mu-p~{\rm results}} \simeq 0.06~{\rm fm}^2.& \nonumber
\end{eqnarray}
The difference between $r_{p,1}$ \cite{CODATA} and $r_{p,3}$ \cite{muH} is $5\sigma$ while the difference between 
$r_{p,2}$ \cite{scattering} and $r_{p,3}$ is 
$4.6\sigma$ (for an up-to-date theoretical analysis see Ref. \cite{Jentschura}). 
Part of this discrepancy may be related to the model dependence of the proton form factor used in various extractions of $r_p$~\cite{noNP-hill}, and it is conceivable that further scrutiny of SM predictions can close this gap~\cite{justSM}.
At the moment, however, this discrepancy 
stands and has stimulated 
investigations of new interactions 
that could potentially be responsible for the difference \cite{Jaeckel,Barger,Itay}. 
The difficulties associated with 
such an enterprise 
stem from the fact that 
the difference (\ref{pattern}) requires 
the strength of the new interactions to be 
on the order of $O(10^4 G_F)$, which is impossible to attain without new light states
below $1$ GeV. 

It is
easy to see that the kinetically mixed vector (\ref{kinmix}) 
cannot explain the observed pattern. 
In the presence of $V$-exchange,  
the inferred $r_p$ would actually depend on the
effective momentum transfer $|q|$ involved \cite{Pospelov}. 
For $r_{p,1}$ ($r_{p,3}$), this corresponds to the inverse Bohr radius
$\sim \alpha m_e$ ($\alpha m_\mu$), while for $r_{p,2}$ the momentum transfer is much larger.
The effect of the extra attraction generated by $V$ will be interpreted as the {\em largest negative} correction to the charge radius 
for the experiment that involves the {\em smallest} $|q|$. 
Therefore, a kinetically mixed vector predicts 
$r_{p,1}< r_{p,3} < r_{p,2}$, which is not consistent with the observed pattern (\ref{pattern}). 
One can easily show that the inclusion of several kinetically mixed vectors does not change this pattern. Another logical possibility is a repulsive Yukawa force between protons and muons/electrons, {\em e.g.}, as may occur 
if there is a new force with gauged baryon number
and kinetic mixing with photons. However, in this case 
the natural pattern will be $r_{p,2}< r_{p,3} < r_{p,1}$, which again disagrees with (\ref{pattern}).

In Refs. \cite{Barger,Itay} a purely phenomenological approach 
to explain
(\ref{pattern}) was taken, in which  
dimension 6 operators 
$(\bar \mu \gamma^\alpha \mu) (\bar p \gamma_\alpha p)$ or $(\bar \mu  \mu) (\bar p p)$
are mediated by the exchange of a new light vector or scalar particle. 
Scalar mediators of this type are reminiscent of 
a very light Higgs boson and 
will face stringent constraints from rare meson decays
and neutron-nucleus scattering \cite{neutron}. 
Vector mediators 
are 
more 
promising, 
but in order to be integrated with the rest of the 
SM, the following conditions must be met: 

\vspace{5pt}

\noindent {\em i.} The interactions must be formulated in terms of SM fermion 
representations,\\
{\em ii.} No new interactions stronger than $G_F$ can exist between neutrinos and 
nucleons or electrons, \\
{\em iii.} No new electrically charged elementary particles with masses below 100 GeV can exist, \\
{\em iv.} The model must have the possibility of a UV completion at or above the weak scale,\\
{\em v.} The model must be consistent with a variety of tests from QED and particle physics in the MeV energy range.

\vspace{5pt}

\noindent The second condition comes from the wealth of data on neutrino scattering in the $E\sim 10$ MeV energy range 
and neutrino oscillations and is emphasized here because it serves as a powerful model discriminator. 
Indeed, the interaction of a new particle $V$ with the lepton vector current 
may be viewed as a subset of the interaction with left- and right-handed SM fermion currents,
\be
V_\alpha \bar l\gamma_\alpha l \; \subset \; V_\alpha( c_1 \bar L \gamma_\alpha L + c_2 \bar R \gamma_\alpha R), ~c_1 \neq -c_2.
\ee
The left-handed fermion doublet $L$ includes a neutrino field, so the requirement {\em ii.}
is equivalent to $c_1=0$. 
This forces 
$V$ to couple to the pure right-handed fermion current. 
The absence of large neutral right-handed currents for electrons follows 
from PNC tests in the electron sector, and we therefore conclude that 
the most promising coupling of a vector particle that can explain (\ref{pattern}) is to the right-handed muon. 
 
{\em Models with gauged $\mu_R$.} We now focus on the class of models 
based on a new $U(1)_R$ gauge symmetry with 
quantized $\mu_R$ number. The Lagrangian is  
\be
\label{Lagr}
{\cal L} =  -\fr{1}{4} V_{\alpha\beta}^2 + |D_\alpha \phi|^2  + \bar \mu_R iD \!\!\!\!/ \,\mu_R -
\fr{\kappa}{2} V_{\alpha\beta} F^{\alpha\beta} -{\cal L}_m.
\ee
Here $V$ is the $U(1)_R$ gauge boson, $\phi$ is a new Higgs field, neutral under the SM gauge group and charged under 
$U(1)_R$, that condenses $\langle \phi \rangle \equiv v_R/\sqrt2$, $D = \partial + i g_R Q_R V+i e Q_{EM} A$, and $\kappa$ is the mixing angle parameter.
The mass term for the muon is necessarily a higher-dimensional operator involving both $\phi$ and 
the SM Higgs field $H_{SM}$ generated at a high scale $\Lambda$,
\be
{\cal L}_m = \bar L \mu_R H_{SM} \frac{\phi}{\Lambda} + {\rm h.c.} \to 
 \bar \mu \mu  
\frac{vv_R}{2 \Lambda},
\label{massL}
\ee
with $v_R/(\sqrt2\Lambda)$ entering as an effective SM-like Yukawa coupling for the muon. As we shall see below, the 
range for $v_R$ suggested by the charge radius phenomenology is fully consistent with $\Lambda$ being at the 
weak scale. Therefore, we are not concerned with building an explicit model that provides a UV completion to 
${\cal L}_m$. The physical excitation of $\phi$ is a new muon-specific Higgs scalar $S$ 
in the mass range $m_S \la v_R$.

The model (\ref{Lagr}) suffers from gauge anomalies involving the photon and $V$. 
It is possible to restore gauge invariance by introducing dynamical scalar `gauge' degrees of freedom. The price for maintaining gauge invariance is that the theory becomes nonrenormalizable, with a UV cutoff  $\Lambda_{UV}$ above which calculability of the theory is lost. The estimate for $\Lambda_{UV}$ may be obtained, {\em e.g.}, from the radiative three loop vector self energy diagram ~\cite{Preskill}:
\be
\Lambda_{UV} \le \frac{(4 \pi)^3}{e g_R^2} m_V \sim 
700 {\rm GeV} \left(\frac{m_V}{10~{\rm MeV}} \right) \left(\fr{g_R}{e}\right)^{-2}.
\label{anomaly}
\ee
We observe that vectors in the range $m_V \sim10-100$ MeV with couplings $g_R \sim 0.01-0.1$ are consistent with a UV cutoff well above the TeV scale.  There are of course examples 
of perturbative cancellations of the
anomalies, such as
quantized $\mu_R+s_R-c_R$. 
Such a scenario faces severe
constraints from quark flavor physics, $c\bar c $ resonance decays, and parity-violating tests 
involving nucleons and appears to be thoroughly excluded. Therefore, we choose the model 
(\ref{Lagr}) as the best candidate to describe new muon-specific forces, which is a consistent effective field theory valid below $\Lambda_{UV}$~(\ref{anomaly}).

{\em Phenomenological constraints.} 
From the Lagrangian (\ref{Lagr}) we obtain the couplings of the new vector and scalar particles
to fermions,
\begin{eqnarray}
\label{couplings}
&g_{V}^\mu = -e\kappa - \displaystyle{\fr{g_R}{2}};~ g_{A}^\mu = -\displaystyle{\fr{g_R}{2}};~ g_{V}^p=-g_V^e = e\kappa 
\\
& g_{A}^{e,p,n}=g_V^n = 0; ~ g_S^\mu = |g_R|m_\mu/m_V, & \nonumber
 \end{eqnarray}
where $e=\sqrt{4\pi \alpha}$ is the positron charge. With the couplings (\ref{couplings}), we calculate the corrections to the energy levels of the ordinary and muonic hydrogen 
that will be interpreted as corrections to the proton charge radius,
\begin{eqnarray}
\left.\Delta r_p^2\right|_{e \rm H} = -\fr{6\kappa^2}{m_V^2};
\left.\Delta r_p^2\right|_{\mu \rm H} = -\fr{6(\kappa^2+\eta)}{m_V^2} f(a m_V)
\end{eqnarray} 
where $a = (\alpha m_\mu m_p)^{-1}(m_\mu +m_p)$ is the $\mu$H Bohr radius, 
$f(\hat x) \equiv \hat x^4(1+\hat x)^{-4}$, and 
$\eta \equiv \kappa g_R/(2e)$.
The difference $\left.\Delta r_p^2\right|_{e \rm H}  -\left.\Delta r_p^2\right|_{\mu \rm H}$  
must be consistent with the observed pattern (\ref{pattern}) 
and requires $\eta$ to be positive. 
In the scaling regime of $am_V \gg 1$ one has
\be
\label{condition}
\fr{\eta}{m_V^2}\simeq \fr{\Delta r^2}{6} \simeq 0.01~{\rm fm}^2 \simeq \fr{2.5\times 10^{-5}}{(10 ~{\rm MeV})^2}.
\ee
In the same regime, the model predicts that future experiments with $\mu$He would detect
the effective charge radius of the helium nucleus shifted down by $\Delta r_{\rm He}^2 = - 0.06$fm$^2$.

Another 
important constraint comes from the measurement of $g-2$ of the muon, which 
currently displays a $\sim 2-4\sigma$ discrepancy with the SM prediction depending on the estimate of the hadronic contribution. 
The individual one-loop contributions from $V$ and $S$ are large ({\em i.e.} 
enhanced compared to the pure vector case \cite{Pospelov} by $m_\mu^2/m_V^2$) and of opposite sign, so that for the choice of parameters (\ref{condition})
they must cancel. 
This mutual cancellation must 
happen at a per-mille level, and should 
be considered as the main phenomenological drawback of the model (\ref{Lagr}). Nevertheless there do exist choices of 
parameters that simultaneously account for the charge radius data and the muon $g-2$  discrepancy.

Other constraints that must be taken into account are the electron $g-2$ determination vs. independent measurements
of $\alpha$ \cite{alpha} and tests of $d-p$ transitions in muonic Si and Mg \cite{SiMg}, for which no deviations from standard QED predictions were found. 
Table I displays three benchmark points for $m_V = 10,~50,~100$ MeV for which all constraints are satisfied. 
Vector masses $m_V \lesssim 10$ MeV are excluded by muonic Si,Mg data and tests of $\alpha$. 
\begin{table}[t]
\begin{center}
\begin{tabular}{c|ccc}
Parameter & ~~Point A~~ & ~~Point B~~ & ~~Point C~~ \\ \hline\hline
$m_V$ & $10$ MeV & 50 MeV & 100 MeV\\ 
$m_S$ & $102.84$ MeV & 90.44 MeV & 84.97 MeV\\ 
$g_R$ & 0.01 & 0.05 & 0.07\\ 
$\kappa$ & 0.0015 & 0.0075 & 0.02\\ 
$\eta$ & $2.5\times 10^{-5}$ & $6.2\times 10^{-4}$ & $2.3\times 10^{-3}$\\ 
$v_R$ & 1 GeV & 1 GeV & 1.4 GeV \\ 
\hline
\end{tabular}
\end{center}
\caption{Benchmark points for the model that pass all phenomenological constraints.} 
\label{table1}
\end{table}

Additional constraints on gauged $\mu_R$ theories
depend on the decay channels of $S$ and $V$. 
If no new states charged under $U(1)_R$ exist below $m_V/2$,  the 
gauge boson $V$ will decay to $e^+e^-$ pairs and thus be subject to tests at lepton colliders and fixed target experiments~\cite{eefixed}. In particular, a preliminary search for the rare decay mode $\phi \rightarrow \eta V$  would disfavor models with 
$\kappa\sim O(10^{-2})$ and $m_V$ above 30 MeV~\cite{kloe}. If new decay channels for $V$ are allowed these bounds  
can be relaxed. 
Among model independent probes, the different couplings of $V$ to muons vs. electrons (\ref{couplings}) suggest 
nonuniversal leptonic branchings of $J/\psi$ and other narrow vector resonances. 
Current data \cite{Jpsi} is only sensitive to $\eta \ge O(10^{-2})$, which does not 
probe the most interesting $m_V \la 100$ MeV regime. An alternative way to search for 
$V$-exchange is to study the $O(\eta)$ forward-backward asymmetry in $e^+e^-\to \mu^+\mu^-$ annihilation at medium energy high-luminosity facilities with longitudinally polarized beams. Finally, since the gauge boson couples preferentially to $\mu_R$, there is no reason for  charged lepton flavor conservation. 
One must therefore make some assumptions about the underlying structure of the Yukawa matrices, implying new flavor-related physics in the UV.

\begin{figure}
\centering
\includegraphics[width=0.34\textwidth]{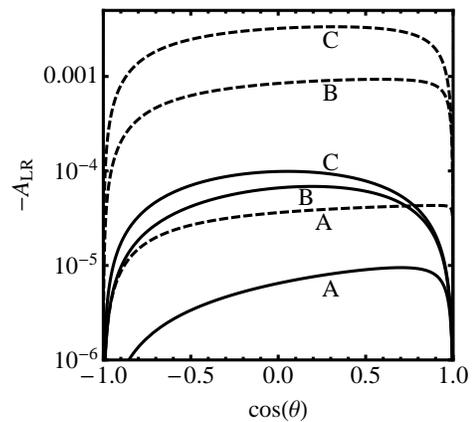}
\caption{The asymmetry $A_{\rm LR}(\theta)$ defined in Eq.~(\ref{asymmetry}) for the benchmark points labeled A, B, and C in Table \ref{table1}.  The solid curves are for $p=29~{\rm MeV}/c$ and dashed curves for $p=200~{\rm MeV}/c$.
}
\label{figALR}
\end{figure}

{\em New parity-violating effects.} Despite the existence of 
polarized muon sources, no tests of PNC 
in neutral currents involving low-energy muon beams have been performed. This is because the maximum muon intensity 
corresponds to $p=29~{\rm MeV}/c$, where the parity violating asymmetry due to the weak interactions will not 
exceed $O(10^{-7})$. With the introduction of a new vector force coupled to $\mu_R$, 
the PNC effects are greatly enhanced. For the scattering of  semi-relativstic muons on a
heavy nuclear target, the asymmetry is given by 
\be
\label{asymmetry}
A_{\rm LR} = \fr{d\sigma_{\rm L}-d\sigma_{\rm R}}{d\sigma_{\rm L}+d\sigma_{\rm R}}\simeq -\eta \beta \fr{Q^2}{Q^2+m_V^2}\fr{1 + \cos(\theta)}{1-\beta^2\sin^2(\theta/2)},
\ee
where $Q$ is the momentum transfer of the elastic scattering, $Q^2/p^2 = 4\sin^2(\theta/2)$, 
$\beta = |p|/E$, and $\rm L(R)$ label the incoming muon's helicity. 
Notice that the same combination of couplings $\eta$ governing the correction to $r_p$  
also determines the asymmetry. The PNC asymmetry $A_{\rm LR}(\theta)$ 
is presented in Fig.~(\ref{figALR}) for the three benchmark points in Table \ref{table1} with
two reference values of incident muon momentum, $29$ and $200$ MeV. The asymmetry can vary in a broad range from 
$10^{-5}$ to $10^{-3}$, becoming larger for $Q^2 > m_V^2$ due to the scaling relation  (\ref{condition}).

We next investigate the feasibility of achieving statistical sensitivity to $A_{LR}$ in 
Eq.~(\ref{asymmetry})
in a Rutherford-type scattering setup. 
Since low-energy muons are easily stopped, 
counting rates are maximized with the use of high $Z$ thin foil targets,
while the optimal $Z$ should be determined from the combined analysis of 
statistical and systematic errors. If, for example, a tungsten ($Z=74$) foil of 
$d=0.01$ mm thickness is used, the muons will lose only  $\sim 5\%$ 
of their kinetic energy. 
Assuming a muon-counting detector with 
full azimuthal coverage and 
polar angle coverage in the range from 60 to 80 degrees
where the asymmetry is maximized, one 
obtains
the following probability for the scattering of a muon at a large angle:
\be
P = d \times N_{\rm atoms}\times {\rm Volume}^{-1} \times \overline{ \sigma_{\rm Rth}} \sim 6\times 10^{-4}.
\ee
With this probability, the time required to 
collect $N \sim (A_{LR})^{-2}$ events is given by 
\be
t|_{N\sim 10^8} = \fr{N}{P\Phi_\mu} \sim 1600~{\rm s} \times \fr{10^8~{\rm muons/s}}{\Phi_\mu},
\ee
where we have normalized the muon flux to the highest modern beam intensities 
\cite{beam}. It is thus apparent that 
the statistical uncertainty will not be a limiting factor 
in detecting
parity violating asymmetries of order $10^{-4}$.

Another promising avenue in the search for anomalous PNC effects 
is the study of parity-violating decays of $2s$ states in muonic atoms, in which PNC 
will manifest in the enhanced one-photon rate of $2s$ decays. 
To illustrate, we assume 
$\eta/m_V^2$ is
fixed by the scaling limit (\ref{condition}) and compute the $2s_{1/2}$-$2p_{1/2}$
mixing in $\mu^4$He.
The results for the mixing angle $\delta$, ratio of $E1$ to $M1$ 
amplitudes for the one-photon decay of the $2s_{1/2}$ state,
and rate for the one-photon decay are given by
\begin{eqnarray}
\delta \simeq 3\times 10^{-5};~~ \fr{A(E1)}{A(M1)} \simeq 60;~~~~~~~~~~~~\\\nonumber
\Gamma^{\gamma}_{2s\to 1s} \simeq  1.9\times 10^3~{\rm Hz};~~
\Gamma^{\gamma}_{2s\to 1s}/\Gamma^{\gamma\gamma}_{2s\to 1s}\simeq 0.018.
\end{eqnarray}
The rate of the one-photon decay is only marginally smaller than the 
one-photon quenching rate \cite{gas} at a gas pressure of 4hPa, at which the
$\mu^4$He
Lamb shift experiment is planned, and the presence of such decays can
be searched for at different gas pressures 
(with a modest improvement of the $2\gamma$ background rejection). 

Finally, despite the absence of parity-violating couplings to the electron at tree level,  
an $e$-nucleus parity-violating amplitude will still occur at the two loop level.
Given the accuracy achieved in tests of PNC with electrons, it is therefore highly desirable to 
calculate this effect,
which may lead to an 
independent constraint on gauged $\mu_R$ models.

To conclude, we have argued that the class of models with gauged $\mu_R$ represents one of 
the few 
possibilities in which the discrepancy between 
$e-p$ and $\mu-p$ determinations of the proton charge radius 
can be reconciled with new physics. 
Although anomalous, these models constitute valid effective field theories which can in principle be UV-completed at the weak scale.
The simultaneous explanation of
the $r_p$ data and muon $g-2$ discrepancy requires a tight correlation
between the scalar and vector masses. 
A striking consequence of this class of models is the existence of 
enhanced PNC effects in the muon sector that can be searched for at
existing  muon beam facilities. 

We would like to thank Drs. A. Antognini, C. Burgess, A. Czarnecki, A. de Gouv$\hat{\rm e}$a, K. Kirch,  W. Marciano, G. Marshall, A. Olin, M. Roney and I. Yavin for helpful discussions. 

\end{document}